# Mapping Organization Knowledge Network and Social Media Based Reputation Management

Andry Alamsyah, Maribella Syawiluna

*School of Economic and Business, Telkom University*
*Bandung, Indonesia*

andrya@telkomuniversity.ac.id

**Abstract**

Knowledge management are important aspects in an organization, especially in ICT industry. Having more control of it is essentials for the organization to stay competitive in the business. One way to assess the organization knowledge capital is by measuring employee knowledge network and their personal reputation in social media. Using this measurement, we see how employee build relationship around their peer networks or clients virtually. We also able to see how knowledge network support organization performance.

The research objective is to map knowledge network and reputation formulation in order to fully understand how knowledge flow and whether employee reputation have higher degree of influence in organization knowledge network. We particularly develop formulas to measure knowledge network and personal reputation based on their social media activities. As case study, we pick an Indonesian ICT company which actively build their business around their employee peer knowledge outside the company.

For knowledge network, we perform data collection by conducting interviews. For reputation management, we collect data from several popular social media. We base our work on Social Network Analysis (SNA) methodology. The result shows that employees knowledge is directly proportional with their reputation, but there are different reputations level on different social media observed in this research.

## I. Introduction

Knowledge management (KM) is a substantial aspect of an organization. It proposed to improve decision-making, responsiveness to customer, efficiency of work, productivity, and ability to innovate [1]. Furthermore, human resources become an essential factor to make a good reputation for organization, since they are the actors who create, execute and develop organization's plans and strategies. Therefore, they can influence the organizations' reputation, whether they will create better or worse reputation for the organization. It depends on the level of their knowledge and how they utilize their knowledge in the organization. In another word, if an organization have more human resources who have a high level of knowledge, that organization will have a better reputation. Moreover, the organization can also increase its level of competitiveness [2]. Personal knowledge network (PKN) within the organization is an alternative to model overall organization KM [3].

Organizations' reputation is influenced by personal reputation within the organization [4]. Nonetheless, it is difficult to measure the reputation based on the relationship of each actor. At first, developing a network which includes all the actors takes too much effort in terms of time and money. For example, if the organization wants to get some information about their employee, they need to stay close to the employee, to monitor, or they can collect the data from their employee directly. This way, the organization needs to observe or analyze the data one by one.

Nowadays, the technology development permit increase usage of social media. People start to change their behavior from conventional such as offline activities to digital recorded activities. The collaboration of technology, internet, and social media allows people to communicate, exchange information and express themselves, which leads to a network creation and alters the relationship between people. Organizations can collect this digital trace in the form of massive data then analyze it to measure personal reputation.

According to [5], the most popular social media are *Facebook*, *Twitter*, *Instagram*, and *LinkedIn*. The problem raises since the formula to get the information about reputation from those social media is not available yet. Therefore, this study aims to construct formulas of KM and reputation for network modeling. The formulas are developed to measure KM and personal reputation based on their social media activities. In this research, as case study, we pick an Indonesian ICT company which actively build their business around their employee peer knowledge outside the company. There are 48 core employees in the company who become our respondent. These employees represent the higher management position in the company.

## II. LITERATURE REVIEW

### *A. Knowledge Management*

Knowledge Management (KM) is an activity which has a systematic approach to capture, structure, manage, and disseminate knowledge through an organization [6]. Human, technology, organization, and management process are the critical success factors of KM in practice. Previous studies highlight that KM which is the part of human capital plays important role in the competitiveness of large firms and small and medium enterprises [7][8]. In today information era, company valued by their human capital especially in ICT industry. Therefore, KM becomes one of the most important aspect related to the sustainability of the organization. There is also effort to map KM using Big Data and Social Network Analysis (SNA), which clearly a new approach to the KM field [3][11][12].

In general, KM process is divided into three phases. First, the creation phase which explains about acquiring and validating knowledge [8]. Second, the storage phase which explains about retaining and organizing knowledge. The last phase is transfer phase. In this phase, knowledge is exchanged and shared by several actors.

### *B. Personal Reputation*

Personal reputation explains social network of some individuals by reporting the information about people within individual life and their relationship [9]. That personal reputation was given by considering other people's judgment toward individuals' activities. It seems like accumulating other people's perception in long-term about what individual has been done in their life.

Generally, reputation has three dimensional scales [10]:

#### *1) Task Reputation*

Task reputation explains the personal reputation based on their ability to finish tasks [9]. In another word, in the context of the organization, task reputation has a close relationship with individuals' performance. Individuals have high possibility to be famous and have a good reputation in their organization if they have high ability to accomplish their tasks. For an instance, an individual is judged as an expert in some area if they have consistent performance in that area. Furthermore, they will be expected to finish similar job in the future. It indicates that those individual's successes to build a good reputation in their social life.

A wide range of literature developed several indicators to measure task reputation. Individuals who havehigh level of task reputation has characteristics such as:

a) An expert in their area
b) *Often asked for advice regarding work-related issues*
c) Often asked to solve problem regarding technical issues at work
d) Have good understanding about technical systems of the workplace

*2) Social Reputation*

As a social human being, an individual is created to be social units which cooperate to achieve their goal. Therefore, they often create a relationship with other people within or outside their group. A successes or failures to create relationship depends on their reputation in social life.

The fundamental of social reputation is a norm. The social reputation of individuals depends on their ability and willingness to live under control from the social norm [8]. It does not need to support their activities related to their job performance. Meanwhile, it will guide and help individuals to interact in their social life.

There are several characteristics of individual who have a good social reputation, such as:

a) Has attractiveness to make other people feel comfortable
b) Often involved in social events
c) Well-liked by others
d) Popular in their social life

*3) Integrity Reputation*

Reputation is a tool to predict personal behavior and the integrity is one of the most important components to predict it. Integrity reputation can be seen in decision-makingprocess of any situation. Individuals who have a better reputation in any area does not mean they also have a better integrity reputation [9]. It's because each reputation scale has their own parameters or own needs. Different from task reputation and social reputation, characteristics of individuals who have good integrity reputation are:

a) Seen as a person who has high integrity
b) Known for being upstanding person
c) Can be trusted
d) Has high moral character

## III. METHODOLOGY

### *A. Case Selection*

The case study of this research is a small startup ICT company focusing on social media monitoring and analyzing. As an emerging business in big data analysis which develops their relationship with customers through social media, virtual network becomes an essential business aspects. Therefore, individual's reputation in virtual world within the organization must be measured. However, knowledge transfering from past employee does not work well in this organization. Moreover, the organization does not have a good record of data such as employee information. This organization has only recorded employee' data as recent as 2016, although it was established in 2010.

### *B. Data Collection*

Data collection is performed by conducting an interview and handing questionnaire. The interviewees are the organizations' employees. The interview uses task reputation approach. The questionnaire format is in Fig. 1., while rank table is in Fig. 2. Basically, this questionnaire asks each employee who they regard as a knowledgeable colleague, they rank from the best to the worst.

**1. Profil Responden**

Nama Responden :____________________

Jabatan :____________________

**2. Ranking**

**Keterangan :**

**Knowledge** yang dimaksud pada penelitian ini adalah**pengetahuan** yang dimiliki seseorang yang dapat **mendukung kegiatan di lingkup pekerjaannya**.Knowledge dalam konteks ini mengacu pada pengetahuan seputar Perusahaan Mediawave. pengetahuan tersebut dapat berupa **knowledge tentang perusahaan** dan **skill** yang dimiliki karyawan **untuk menyelesaikan pekerjaannya dan menunjang jalannya perusahaan.**

**Petunjuk pengisian kuesioner:**

Pada halaman selanjutnya terdapat tabel karyawan PT Media Wave Interaktif.

**Pilihlah 10 karyawan** yang menurut anda **memiliki knowledge dan berpengaruh pada perusahaan**. Berikan ranking pada kesepuluh karyawan tersebut

**Ranking tertinggi** dimulai dari angka 1 dan selanjutnya mengikuti dengan urut hingga ranking 10.

Fig. 1. Questionnaire Format

This study also performs data crawling from employees' social media account. This study analyzes data from *Instagram*, *Twitter*, and *LinkedIn*. The reason to choose those is come from survey result, most of the employees only use those three social media. The attributes used in the data crawling processes are the number of post, the number of followers and the number of following. The data that have been collected were organized into a dataset to create a network of individuals' KM and reputation.

**Tabel Ranking Pengetahuan Karyawan**

| NO | NAMA | Ranking |
|---|---|---|
| 1 | YOSE RIZAL | |
| 2 | ERIK PALUPI | |
| 4 | ADI SETIADI | |
| 5 | IWAN PRASETIA | |
| 6 | AHMAD MIFTAHUDDIN | |
| 7 | HADRIAN FEBRIYANSAH | |
| 9 | JEJEN ZAENAL MUTAQIN | |
| 10 | TANTI SETIAWATI | |
| 11 | ACEP AHMAN HIDAYAT | |
| 12 | FINA AGUSTINA | |
| 13 | MUHAMMAD SANNY | |
| 14 | AYU LAKHSMI DAENG UGI | |
| 15 | ARIP SAMSUL MAARIP | |
| 16 | SOFIA DEWI | |
| 17 | ARIEL GARDAN | |
| 18 | ALFANO ANDARU YUSUF | |
| 19 | ARIF WICAKSONO | |
| 20 | ROSMANSYAH | |
| 22 | ANISA BELLA FATHIA | |
| 23 | NUR ASTI TUNJUNG SARI | |
| 24 | RULI SETIAWAN | |
| 25 | GANJAR GHANI HILMAN | |
| 26 | ANIS HAIFA | |
| 27 | FIFI FARIDA | |
| 28 | NADIA SHABILLA | |
| 29 | RYAN ARIESTA | |
| 30 | GITA KRISTA | |
| 31 | EKA RAFI DIMASYONO | |
| 32 | AMAJIDA TAMIMI | |
| 33 | ZALDY RIO VALDERAMA | |
| 34 | PRIMANOLA PERDANANTI | |
| 35 | RENI RAHMAWATI | |
| 36 | ELIN HERLINA | |
| 37 | MOCHAMAD LUTFI | |
| 38 | ALDO BANGGA PERMANA | |

| NO | NAMA | Ranking |
|---|---|---|
| 39 | M. RAIZA RONALDO | |
| 40 | INTAN PARIZTYAN E. | |
| 41 | HAIDIR MAPPALEWA | |
| 42 | PHANIE SITI FAUZIAH | |
| 43 | TUBAGUS RAKA BAGJA | |
| 44 | ITHO SURYOPUTRO | |
| 45 | TASYA IMELAWATI | |
| 46 | SONY BANYU ARGA | |
| 47 | ILHAM MAULANA | |
| 48 | RAHMANDA MARYUDIE | |

*- Terima kasih atas perhatian dan kerjasama -*

Fig. 2. Knowledge Ranking Table

### C. Network Methodology

We construct knowledge network based on Social Network Analysis (SNA) [11] methodology from the questionnaire data. The network model will be undirected network which capture the reality on how an employee vote their colleagues. The edge weight has also applied to give some importance measurement of voting. For example, employee A vote employee B in higher rank than employee C, so we have undirected weight edge A to B higher than A to C.

We use centrality metric to measure the most influential employee in the organization. Because of directed network nature, we have in-degree and out-degree measurement. In-degree measure how many people vote the actual employee, while out-degree measure how many votes has been given by an employee.

### D. Formula Reputation Construction

After studying some previous researches related to the reputation and knowledge network based on SNA, we found that the specific formula to analyze personal reputation in social media using SNA has not developed yet. Therefore, this study tries to develop personal reputation formula to calculate personal reputation on *Twitter, Instagram*, and *LinkedIn* using SNA methodology. The formula was developed by considering social reputation approach and 11 elements of popularity from the previous researches [2][9][10]. Then, expert's judgments and comments were also considered to validate the formula.

#### 1) Formula of Personal Reputation on Twitter and Instagram

Formula to measure personal reputation ($R$) on *Twitter* and *Instagram* were developed based on *Zinko* principle on reputation [10]. This formula is built by comparing the number of followers with the number of post and number of following. The formula (1) shows that someone with a good personal reputation is someone who has more followers but less number of post and less number of following. In another word, this individual has good attractiveness to make other people follow them. However, the attractiveness does not fully depend on their high frequent to post something in their social media.

$$R = \frac{\sum follower}{\sum post} + \frac{\sum follower}{\sum following} \tag{1}$$

#### 2) Formula of Personal Reputation on Linkedin

*LinkedIn* is one of social network services that connects peoples in their professionals life. Therefore, *LinkedIn* components can be used to predict personal reputation in their professional area. Personal reputation in *LinkedIn* ($Rl$) is calculated by comparing the ratio number of endorsements and the number of connection as the first element, we also add ratio of the number of endorsements and the number of skill as the second additive element. The formula (2) indicates that the ratio of endorsements comparing to connections and skills give representatives personal reputations.

$$Rl = \frac{\sum endorsements}{\sum connections} + \frac{\sum endorsements}{\sum skills} \tag{2}$$

## IV. RESULT AND ANALYSIS

### *A. Modelling The Knowledge Network*

First, we analyze employee questionnaire data by constructing knowledge network based on SNA. The map result shown in Fig. 3. It describe the modeling of mapping employees' knowledge network in organization. The result consists of nodes which symbolized knowledge level of each employee compared to the other employee. Those nodes are calculated based on accumulated score given by their peer colleagues, thus it shows knowledge level ranking. Bigger size of node indicates higher position of knowledge level. The result also consists of directed edges. The edges are mapping the relationship, in this case knowledge voting among employees. Weighted degree shows their accumulated score. Overall, the degree of centrality gives an information about employees who have the best knowledge level in the organization according to their colleagues. There are three employees who have highest degree of centrality, they are employee number 1, 2, and 16 (shown in Fig. 4).

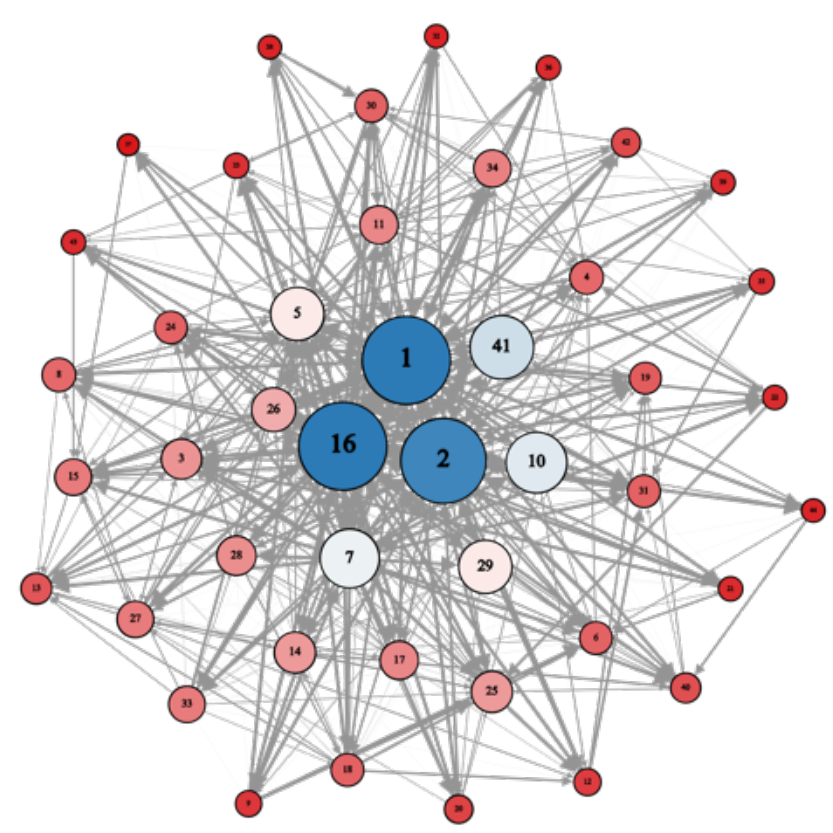

Fig. 3. Knowledge Network

| Nodes | Degree Centrality | |
|---|---|---|
| | Degree | Weighted Degree |
| 1 | 82 | 1828 |
| 16 | 82 | 1786 |
| 2 | 79 | 1343 |
| 7 | 51 | 1011 |
| 5 | 45 | 676 |

Highest knowledg e level

*Fig. 4. Summary of Degree Centrality for Highest Knowledge Level*

### *B. Knowledge Management and Reputation on Social Media*

The second step is crawling data from each personal social media employee accounts from the duration of June 1st – July 1st, 2017. We then apply personal reputation formulas (1) and (2) for each social media. We combine KM based on knowledge network model in Fig. 3. and personal reputation measurement from all social media data (i.e. *Instagram, Twitter,* and *LinkedIn*). Fig. 5. shows the visualization result of the process. We call this process as building KM and reputation on social media network.

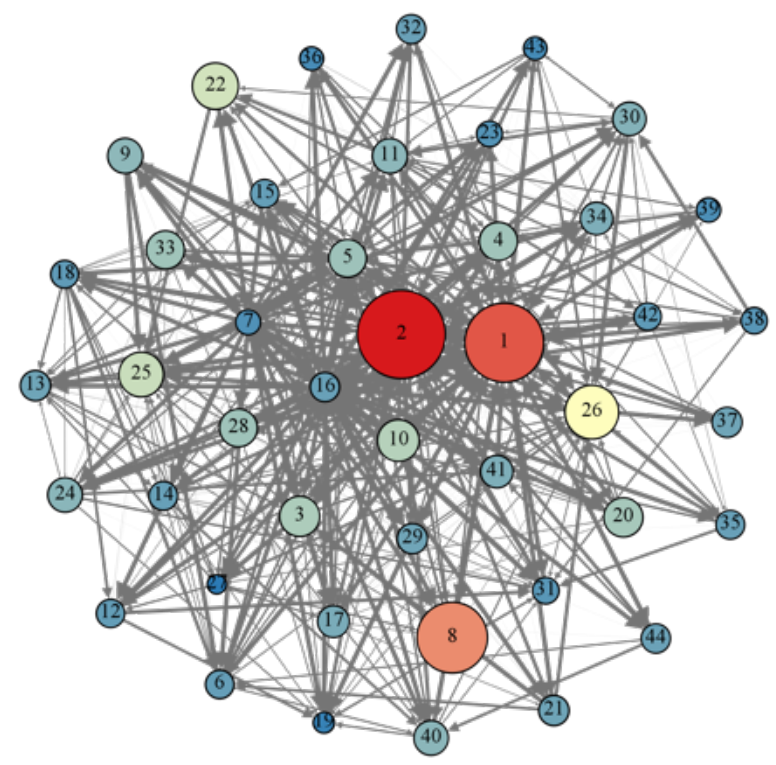

*Fig. 5 Knowledge and Reputation on All Social Media Network*

From Fig. 5., we compare the three biggest nodes with the previous result in 4, where we have slightly different result. Now, we have nodes number 1, 2 and 8 as the three biggest nodes. Node 8 which previously unknown, now become one of the prominent employee based on this rank. Although node 8 have smaller number connection inside the company comparing to node 16 which is the rank number 3 in Fig. 4. Internal connection symbolized by degree centrality no longer affects the size of the nodes, but the actor reputation in social media is. Table 1. shows the result of overall best 5 actors based on the highest reputation network

TABLE 1 SUMMARY OF HIGHEST REPUTATION IN ALL SOCIAL MEDIA

| Nodes | Total Reputation |
|---|---|
| 2 | 1.567 |
| 1 | 1.358 |
| 8 | 1.172 |
| 26 | 0.790 |
| 22 | 0.619 |

In order to have comparable result of network reputation between all social media in this research, we normalize reputation value in a range between 0 and 1 value. The value indicate that someone has a better reputation if the result of calculation is closer to 1. Table 1 gives an information about total employees' reputation based on 3 social media in this research. We found that there are three employees who have the highest reputation in social media. To give a deeper analysis, this study also develops a model for each social media separately. Modeling result of each social media is shown in Fig. 6, 7, and 8. Then the summary of each result is explained in Table 2.

The result shows that each employee has different KM and personal reputation level on different social media. It depends on their activities on social media and how active they are to maintain their reputation on their social media. Other than centrality metric, we also use other network metric that beneficial to the analysis of the KM and reputation network. One of them is density metric.

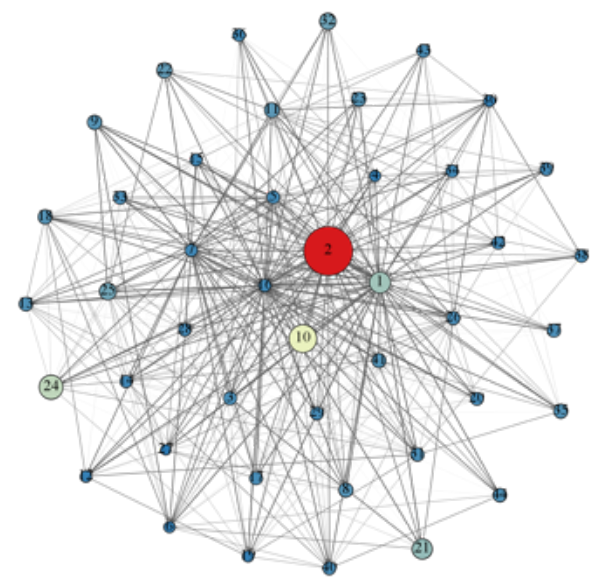

*Fig. 6. Modeling Result of Personal Reputation on Twitter Network*

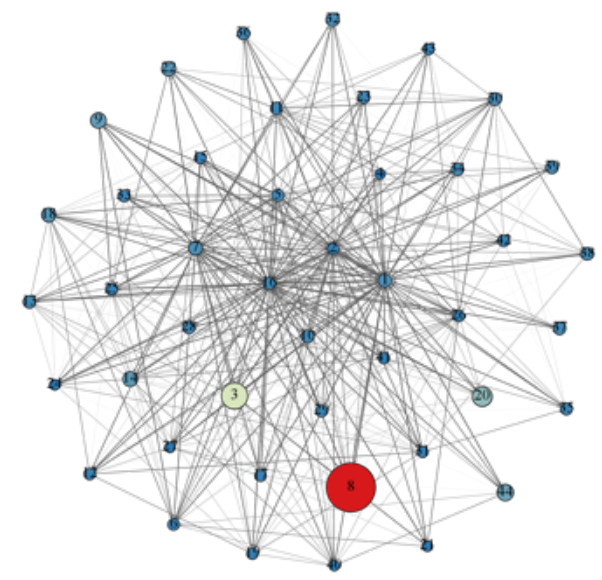

*Fig. 7. Modeling Result of Personal Reputation on Instagram Network*

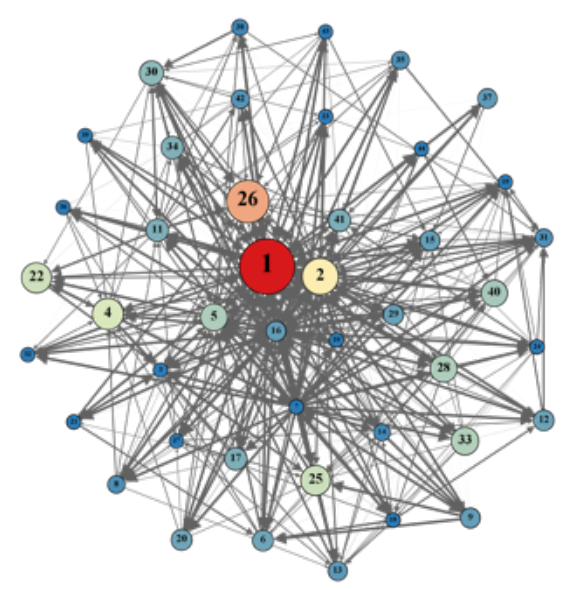

*Fig. 8. Modeling Result of Personal Reputation on LinkedIn Network*

TABLE 2 SUMMARY OF MODELLING RESULT OF KNOWLEDGE MANAGEMENT AND REPUTATION ON SOCIAL MEDIA

| **Reputation on Twitter** | | **Reputation on Instagram** | | **Reputation on Linkedin** | |
|---|---|---|---|---|---|
| **Nodes** | **Reputation** | **Nodes** | **Reputation** | **Nodes** | **Reputation** |
| 2 | 1.000 | 8 | 1.000 | 1 | 1.000 |
| 10 | 0.452 | 3 | 0.409 | 26 | 0.690 |
| 24 | 0.352 | 20 | 0.241 | 2 | 0.536 |
| 1 | 0.280 | 44 | 0.154 | 4 | 0.414 |
| 21 | 0.251 | 9 | 0.130 | 22 | 0.385 |

The density of knowledge network that occur in the entire KM and reputation is 0.658. This implies that the network has high enough density comparing to other social network which commonly have around under the value of 0.2 based on scale free or power law distributions [12][13]. The high density of the network is influenced by a high interaction between actors, the interaction mentioned in this research is ranking activity among employees. This indicates that employees within the organization acknowledge their each colleagues benefit and limitation. It is a positive sign in a company since it can support knowledge distribution process and information easily.

## V. CONCLUSION

This study presents formulas to measure the level of knowledge management and reputation on social media using Social Network Analysis methodology. The results show knowledge network model that support both knowledge management and reputation of the employee. The main components of formula consider the number of posts, the number of followers, and the number of following which describes their activities on their social media. The modeling result shows that individuals who have better knowledge, also have better personal reputation. Meanwhile, their level of personal reputation in a different social media might be different. It depends on their activities and their relationship with other people on social media.

By having this tool as the measurement base, organization can get better understanding on how their employee perform in their peer network. Organization can also map whether their employee is over-performed or under-performed by comparing for example with their salary or other variables. This study also shows that knowledge management and reputation network mapping can outline employee knowledge, predict the optimal knowledge dissemination scenario, which then useful for overall decision support system in the organization and organization competitiveness.

For the future research, we suggest more responden, which means more actors and more relationships to be included. The objective is to understand whether the network complexity can reveal the real organization knowledge network flow problems. We also suggest comparison knowledge network between different organizations, to see which organization have better knowledge network and whether the advantage can be reflected to their performance.

.